\documentclass{article}

\usepackage{PRIMEarxiv}

\usepackage[utf8]{inputenc} 
\usepackage[T1]{fontenc}    
\usepackage{hyperref}       
\usepackage{url}            
\usepackage{booktabs}       
\usepackage{amsmath}
\usepackage{amsthm}
\usepackage{amssymb}
\usepackage{amsfonts}
\usepackage{mathtools}
\usepackage{amsfonts}       
\usepackage{nicefrac}       
\usepackage{microtype}      
\usepackage{fancyhdr}       
\usepackage{graphicx}       
\usepackage{times}
\usepackage{subfig}
\usepackage{xcolor}
\usepackage{array}
\usepackage{cite}
\usepackage{multirow}
\usepackage{adjustbox}
\usepackage{fontawesome}
\usepackage{algorithm,algorithmic}
\usepackage[font=footnotesize,labelfont=bf]{caption}
\usepackage{tikz-network}
\tikzstyle{block} = [draw, fill=blue!10, rectangle, 
    minimum height=1cm, minimum width=1cm]
\tikzstyle{sum} = [draw, fill=black!20, circle, node distance=1.5cm]
\tikzstyle{input} = [coordinate]
\tikzstyle{output} = [coordinate]
\tikzstyle{pinstyle} = [pin edge={to-,thin,black}]
\usepackage{enumitem}

\title{
An Information-Based Micro-Kalman Filter for Satellite Tracking: A Comparative Study
}

\author{
  Moh Kamalul Wafi \\
  Department of Engineering Physics \\
  Institut Teknologi Sepuluh Nopember (ITS) \\
  Surabaya, Indonesia\\
  \texttt{\{kamalul.wafi\}@its.ac.id} \\
}

\begin{document}
\maketitle

\begin{abstract}
Satellite dynamics and tracking remain important challenges in the context of space exploration and communication systems. Accurate state estimation is essential to maintain reliable orbital motion and system performance. 
This paper presents a mathematical framework for satellite state estimation based on a linearized model described by radial and angular states. The model incorporates two types of measurement noise corresponding to range and scaled angular deviations, which are assumed to be mutually independent with known covariance structures. The estimation problem is formulated using the Kalman filter, together with the associated Algebraic Riccati Equation (ARE), leading to both time-varying and steady-state solutions.
In addition, a micro-Kalman filter ($\mu$KF) formulation is considered and compared with the classical Kalman filter, as well as with the extended Kalman filter (EKF), unscented Kalman filter (UKF), and an adaptive Kalman filter under a unified simulation setup.
The results demonstrate that the proposed $\mu$KF achieves estimation performance nearly identical to that of the classical Kalman filter and its variants, with small and bounded estimation errors. The mean square estimation error (MSEE) remains low for all state variables under both noise configurations, confirming the effectiveness of the proposed approach for linear Gaussian systems.
\end{abstract}
\allowdisplaybreaks


\section{Introduction}
Tracking an object is becoming more challenging and it has been studying to get the precise position while tracking it. The object refers to the satellite and it has increasingly attained as one of the most challenging topics due to the attraction of elaborating the outer space. The proposed concept of doing it is to use Kalman filter as conducted by \cite{b1} and \cite{b2} which presented the reduced order of the Riccati differential equation, the tractable of the object in terms of mathematical model, and the the ease in the real implementation in turn. This also stimulates to upgrade the classic Kalman filter in order to obtain another best estimation method as done by \cite{b3}, comprising the upgrade of gradient decent in terms of error covariance. The classic Kalman filter \cite{b4} is emerged so as to compare the method in \cite{b7} in terms of the mean square estimation error (MSEE) along with its average over certain iterations.

The basic concept of the classic Kalman is set from \cite{b4} while the elaborating of the basic is well-presented in \cite{b6} saying the various possibility of any engineering and non-engineering being perturbed by any noises including tracking a moving object. The initial foundation has been inspired by \cite{b9}, \cite{b4} along with \cite{b5} for the concept of filtering.

The classic is matched with the \textit{Micro}-Kalman Filter developed by \cite{b7} which then was upgraded in the following paper presented in \cite{b8}. This algorithm is the beginning of the possibility of Distributed Kalman Filter (DKF). This paper proposes the algorithm compared to the classic in terms of the mean square error and overall performance of the states.  

\section{Problem Formulation}

This paper considers the problem of satellite state estimation based on a mathematical model of orbital motion, as illustrated in Fig.~1. The objective is to accurately estimate the system states under the presence of measurement uncertainties.

The formulation begins with the derivation of a suitable state-space representation corresponding to a nominally circular orbit. Small deviations from this nominal motion are introduced and used to define the state variables, allowing the dynamics to be expressed in a linearized form.

These deviations are assumed to be observed from ground-based measurements, leading to two types of noisy observations. In particular, the measurements correspond to the range deviation from the nominal radius $R$ and the scaled angular deviation from the nominal motion $\omega t$, both affected by independent noise processes with known covariance structures.

Based on this model, the state estimation problem is addressed using Kalman filtering techniques. The estimation is performed under different noise configurations, and the performance is evaluated in terms of the mean square estimation error (MSEE).

\section{Mathematical Model}

Based on the defined objectives, the formulation of this work begins with a state-space representation of satellite orbital dynamics. The model incorporates small deviations from a nominal circular orbit and serves as the foundation for the filtering framework developed in this paper.

\subsection{State-Space Representation}

Consider a satellite of unit mass moving under the influence of a central gravitational force. The motion is governed by Newton’s law under an inverse-square force field. Let $r(t)$ denote the radial distance of the satellite from the center of attraction and $\theta(t)$ denote its angular position at time $t$. The radial motion is described by
\begin{equation}
    \ddot{r}(t) = r(t)\dot{\theta}^2(t) - \frac{G}{r^2(t)}
    \label{eq:radial_dynamics}
\end{equation}
where $G$ represents the gravitational parameter of the system. The terms $\dot{r}(t)$ and $\ddot{r}(t)$ correspond to the radial velocity and radial acceleration, respectively, while $\dot{\theta}(t)$ denotes the angular velocity.

For an ideal circular orbit with constant radius $R$, the angular motion is uniform and can be expressed as $\theta(t) = \omega t$, where $\omega$ is the constant angular velocity. Under this condition, the gravitational parameter satisfies the relation $G = R^3 \omega^2$. The angular dynamics of the system are then given by
\begin{equation}
    \ddot{\theta}(t) = -2 \dot{\theta}(t)\frac{\dot{r}(t)}{r(t)}
    \label{eq:angular_dynamics}
\end{equation}

To obtain a tractable estimation model, the nonlinear dynamics are linearized around the nominal circular orbit defined by $r(t) = R$ and $\theta(t) = \omega t$. Small deviations from this nominal trajectory are assumed. The state vector is defined as
\begin{equation}
    x(t) = \left[x_1(t), x_2(t), x_3(t), x_4(t)\right]^\top \in \mathbb{R}^n
    \label{eq:state_vector}
\end{equation}
where the states represent deviations from the nominal motion. Specifically, $x_1(t)$ denotes the radial deviation from the nominal radius $R$, $x_2(t)$ represents the radial velocity, $x_3(t)$ corresponds to the scaled angular deviation, and $x_4(t)$ represents the scaled angular velocity deviation. These quantities are defined as
\begin{subequations}
\begin{align}
    x_1(t) &= r(t) - R \label{eq:state_x1} \\
    x_2(t) &= \dot{r}(t) \label{eq:state_x2} \\
    x_3(t) &= R\big(\theta(t) - \omega t\big) \label{eq:state_x3} \\
    x_4(t) &= R\big(\dot{\theta}(t) - \omega\big) \label{eq:state_x4}
\end{align}
\end{subequations}

The scaling factor $R$ applied to the angular components ensures that all state variables are expressed in consistent physical units, which is advantageous for numerical stability in estimation. Under the assumption of small deviations, the system can be approximated by a linear time-invariant model. In this formulation, process noise is neglected in the system dynamics, and uncertainty is introduced solely through measurement noise, which will be described in the following subsection.

\subsection{Measurement Model and Noise Characterization}

Measurements are collected at discrete sampling times $t_k = kh$, where $k = 1, 2, \ldots, N$ and $h$ denotes the sampling interval. The observation model is defined as
\begin{equation}
    y_k = H x_k + v_k
    \label{eq:measurement_model}
\end{equation}
where $y_k \in \mathbb{R}^2$ is the measurement vector, $x_k = x(t_k)$ is the state vector, and $v_k \in \mathbb{R}^2$ represents the measurement noise.
The measurements consist of the radial deviation and the scaled angular deviation, which can be written as
\begin{equation}
    y_k =
    \begin{bmatrix}
        y_1(t_k) \\
        y_3(t_k)
    \end{bmatrix}
    =
    \begin{bmatrix}
        x_1(t_k) \\
        x_3(t_k)
    \end{bmatrix}
    +
    \begin{bmatrix}
        v_1(t_k) \\
        v_3(t_k)
    \end{bmatrix}
    \label{eq:measurement_vector}
\end{equation}

The measurement noise $v_k = [v_1(t_k), v_3(t_k)]^\top$ is assumed to be zero-mean Gaussian with covariance matrix
\begin{equation}
    \Sigma_v =
    \begin{bmatrix}
        \varphi & 0 \\
        0 & \psi
    \end{bmatrix}
    \label{eq:measurement_cov}
\end{equation}
where $\varphi$ and $\psi$ denote the variances associated with the radial and angular measurements, respectively. The noise components are assumed to be mutually independent and uncorrelated with the system state.

\section{Kalman Estimation}

The objective of state estimation is to infer the system state from noisy and partial observations. In this work, the system is modeled in discrete time at sampling instants $t_k = kh$, where $k = 1, 2, \ldots, N$.

The state-space model is given by
\begin{subequations}
\begin{align}
    x_k &= F x_{k-1} + q_{k-1}
    \label{eq:state_model} \\
    y_k &= H x_k + v_k
    \label{eq:measurement_model_kf}
\end{align}
\end{subequations}
where $x_k \in \mathbb{R}^n$ is the state vector, $y_k \in \mathbb{R}^m$ is the measurement vector, $F$ is the state transition matrix, and $H$ is the observation matrix. The process noise $q_k$ and measurement noise $v_k$ are assumed to be mutually independent, zero-mean Gaussian sequences with covariances $\Sigma_q$ and $\Sigma_v$, respectively.

The initial state $x_0$ is modeled as a Gaussian random variable with mean and covariance given by
\begin{equation}
    \mathbb{E}[x_0] = \hat{x}_{0|0}, \qquad \mathbb{E}\left[(x_0 - \hat{x}_{0|0})(x_0 - \hat{x}_{0|0})^\top\right] = P_{0|0}
    \label{eq:initial_state}
\end{equation}

The Kalman filter provides a recursive solution for state estimation through prediction and update steps. The one-step-ahead prediction is given by
\begin{subequations}
\begin{align}
    \hat{x}_{k|k-1} &= F \hat{x}_{k-1|k-1}
    \label{eq:state_prediction} \\
    P_{k|k-1} &= F P_{k-1|k-1} F^\top + \Sigma_q
    \label{eq:cov_prediction}
\end{align}

Upon receiving the measurement $y_k$, the estimate is updated as
\begin{align}
    K_k &= P_{k|k-1} H^\top \left(H P_{k|k-1} H^\top + \Sigma_v\right)^{-1}
    \label{eq:kalman_gain} \\
    \hat{x}_{k|k} &= \hat{x}_{k|k-1} + K_k \left(y_k - H \hat{x}_{k|k-1}\right)
    \label{eq:state_update} \\
    P_{k|k} &= \left(I - K_k H\right) P_{k|k-1}
    \label{eq:cov_update}
\end{align}
\end{subequations}

The term $y_k - H \hat{x}_{k|k-1}$ is referred to as the innovation, which represents the discrepancy between the observed and predicted measurements. The Kalman gain $K_k$ is computed to minimize the estimation error covariance, resulting in an optimal linear estimator under Gaussian noise assumptions.

\subsection{Steady-State Gain Matrices}

The recursive implementation of the Kalman filter requires the computation of the error covariance matrix $P_{k|k-1}$ and the gain matrix $K_k$ at each time step. In particular, the evaluation of the Kalman gain involves the inversion of the innovation covariance matrix
\begin{equation}
    S_k = H P_{k|k-1} H^\top + \Sigma_v
    \label{eq:innovation_cov}
\end{equation}
which, in general, requires $O(n^3)$ operations for an $n \times n$ matrix. As a result, the computational complexity of the filter increases significantly with the dimension of the system, and repeated online evaluation of this matrix inversion may become computationally burdensome.

For time-invariant systems, it is therefore of practical interest to consider a steady-state formulation in which the matrices $P_{k|k-1}$ and $K_k$ converge to constant values. In such a case, the gain matrix can be computed offline, thereby reducing the online computational burden while maintaining comparable estimation performance after the transient phase.

Starting from the covariance recursion
\begin{subequations}
\begin{align}
    P_{k|k-1} &= F P_{k-1|k-1} F^\top + \Sigma_q
    \label{eq:ss_cov_pred} \\
    P_{k|k} &= P_{k|k-1} - P_{k|k-1} H^\top
    \left(H P_{k|k-1} H^\top + \Sigma_v\right)^{-1}
    H P_{k|k-1}
    \label{eq:ss_cov_update}
\end{align}
it is natural to ask whether the sequence $\{P_{k|k-1}\}$ converges as $k \to \infty$. If such a limit exists, then
\begin{equation}
    P_{k|k-1} \to P, \qquad K_k \to K
    \label{eq:ss_limit}
\end{equation}
\end{subequations}
for some constant matrices $P$ and $K$, since the Kalman gain $K_k$ depends explicitly on $P_{k|k-1}$. In this case, the time-varying Kalman filter is replaced by a steady-state filter with fixed gain.

By substituting the limiting covariance into the Riccati recursion, one obtains the discrete algebraic Riccati equation (DARE)
\begin{subequations}
\begin{equation}
    P = F \left(P - P H^\top \left(H P H^\top + \Sigma_v\right)^{-1} H P \right) F^\top + \Sigma_q
    \label{eq:dare}
\end{equation}
where $P$ represents the steady-state prediction error covariance matrix. Once $P$ is determined, the corresponding steady-state Kalman gain is given by
\begin{equation}
    K = P H^\top \left(H P H^\top + \Sigma_v\right)^{-1}.
    \label{eq:ss_gain}
\end{equation}
\end{subequations}

Using this constant gain, the steady-state Kalman filter may be written in predictor-corrector form as
\begin{subequations}
\begin{align}
    \hat{x}_{k+1|k} &= F \hat{x}_{k|k}
    \label{eq:ss_predictor} \\
    \hat{x}_{k|k} &= \hat{x}_{k|k-1} + K\left(y_k - H \hat{x}_{k|k-1}\right)
    \label{eq:ss_corrector}
\end{align}
\end{subequations}
where the innovation term $e_k = y_k - H \hat{x}_{k|k-1}$ represents the discrepancy between the actual measurement and its one-step-ahead prediction. Equivalently, the predictor can be expressed in the compact form
\begin{subequations}
\begin{equation}
    \hat{x}_{k+1|k} = F \hat{x}_{k|k-1} + F K e_k.
    \label{eq:ss_compact_predictor}
\end{equation}
The associated predicted measurement is then
\begin{equation}
    \hat{y}_{k+1|k} = H \hat{x}_{k+1|k}.
    \label{eq:ss_output_predictor}
\end{equation}
\end{subequations}

The steady-state formulation is attractive because the matrix $K$ is computed only once, thereby reducing the online computational burden. This simplification is especially useful when the system matrices are time-invariant and the noise statistics remain constant. Although the resulting estimator is no longer adaptive in the same way as the full time-varying Kalman filter, it often provides nearly identical performance after the transient phase has decayed.

The existence and uniqueness of a stabilizing positive semidefinite solution to \eqref{eq:dare} depend on standard detectability and stabilizability conditions. In particular, if the pair $(F,H)$ is detectable and the pair $(F,\Sigma_q^{1/2})$ is stabilizable, then the Riccati recursion converges to the unique stabilizing solution of the DARE. Under these conditions, the steady-state gain $K$ ensures that the estimation error dynamics are asymptotically stable. More precisely, the matrix
\begin{equation}
    F - K H
    \label{eq:error_dynamics_matrix}
\end{equation}
must be Schur stable, that is, all of its eigenvalues must lie strictly inside the unit circle. When this condition is satisfied, the steady-state filter yields a bounded estimation error covariance and a stable predictor.

\subsection{Micro-Kalman Filter ($\mu$KF)}

The micro-Kalman filter ($\mu$KF) considered in this work is based on the information form of the Kalman filter, as introduced in \cite{b7, b8}. This formulation is particularly suitable for distributed estimation and provides an alternative perspective to the classical covariance-based approach.

The system dynamics are given by
\begin{subequations}
\begin{align}
    x_{k+1} &= A x_k + B \vartheta_k
    \label{eq:mukf_state} \\
    y_k &= H x_k + \chi_k
    \label{eq:mukf_measurement}
\end{align}
\end{subequations}
where $\vartheta_k$ and $\chi_k$ are zero-mean, mutually independent white Gaussian noise sequences with covariances $\Sigma_q$ and $\Sigma_v$, respectively. The noise processes satisfy
\begin{equation}
    \mathbb{E}[\vartheta_k \vartheta_l^\top] = \Sigma_q \delta_{kl}, \qquad
    \mathbb{E}[\chi_k \chi_l^\top] = \Sigma_v \delta_{kl}
    \label{eq:mukf_noise}
\end{equation}
where $\delta_{kl}$ denotes the Kronecker delta.

In contrast to the classical Kalman filter, which propagates the covariance matrix $P_k$, the information form operates on the inverse covariance. Define the information matrix
\begin{equation}
    S_k = H^\top \Sigma_v^{-1} H
    \label{eq:info_matrix}
\end{equation}
and introduce the matrix
\begin{equation}
    M_k = \left(P_{k|k-1}^{-1} + S_k \right)^{-1}.
    \label{eq:M_matrix}
\end{equation}

The update step of the $\mu$KF can then be written as
\begin{subequations}
\begin{equation}
    \hat{x}_k = \hat{x}_{k|k-1} + M_k \left( H^\top \Sigma_v^{-1} y_k - S_k \hat{x}_{k|k-1} \right),
    \label{eq:mukf_update}
\end{equation}
which can be interpreted as an information-weighted correction of the prior estimate.

By defining the transformed measurement
\begin{equation}
    z_k = H^\top \Sigma_v^{-1} y_k,
    \label{eq:transformed_measurement}
\end{equation}
the update equation can be expressed more compactly as
\begin{equation}
    \hat{x}_k = \hat{x}_{k|k-1} + M_k \left( z_k - S_k \hat{x}_{k|k-1} \right).
    \label{eq:mukf_compact}
\end{equation}

Following the update, the prediction step is given by
\begin{align}
    \hat{x}_{k+1|k} &= A \hat{x}_k
    \label{eq:mukf_predict_state} \\
    P_{k+1|k} &= A M_k A^\top + \Sigma_q.
    \label{eq:mukf_predict_cov}
\end{align}
\end{subequations}

In this formulation, the matrix $M_k$ plays a role analogous to the posterior covariance in the classical Kalman filter. However, the update is expressed directly in terms of information quantities, which avoids explicit computation of the Kalman gain and enables efficient implementation in distributed settings.

Compared to the classical Kalman filter, the $\mu$KF modifies the sequence of operations by performing the update in the information domain prior to state propagation. This results in a structurally different estimator while preserving the underlying Bayesian interpretation of the filtering problem.

\section{Simulation Results}

The simulation is initialized with the parameters $R = 1$, $\omega = 1$, and $G = 1$, and is evaluated over $N = 1000$ discrete time steps. The sampling interval is chosen as $h = 0.01$. 
The initial state is modeled as a Gaussian random variable
\begin{equation}
    x_0 \sim \mathcal{N}(\hat{x}_{0|0}, P_{0|0}),
    \label{eq:init_state}
\end{equation}
with mean $\hat{x}_{0|0} = [0.1,\, 0,\, 0,\, 0]^\top$ and covariance $P_{0|0} = 0.1 I_{4}$.

Based on the linearized dynamics, the continuous-time state equations are given by
\begin{align}\label{eq:dyn}
    \left.\begin{aligned}
    \dot{x}_1 &= x_2  \\
    \dot{x}_2 &= 3\omega^2 x_1 + 2\omega x_4 \\
    \dot{x}_3 &= x_4  \\
    \dot{x}_4 &= -2\omega x_2 
    \end{aligned}\quad\right\} \quad
    \dot{x}(t) = 
    \begin{bmatrix}
        0 & 1 & 0 & 0 \\
        3\omega^2 & 0 & 0 & 2\omega \\
        0 & 0 & 0 & 1 \\
        0 & -2\omega & 0 & 0
    \end{bmatrix} x(t),
\end{align}
Discretizing the system with sampling interval $h = 0.01$ yields the state transition matrix
\begin{equation}
    F =
    \begin{bmatrix}
        1.0001  & 0.0100  & 0 & 0.0001 \\
        0.0300  & 1.0000  & 0 & 0.0200 \\
        0        & -0.0001 & 1 & 0.0100 \\
        -0.0003 & -0.0200 & 0 & 0.9998
    \end{bmatrix}.
    \label{eq:F_matrix}
\end{equation}

The measurement model follows the formulation introduced in Section~III, where the observation matrix is given by
\begin{equation*}
    H =
    \begin{bmatrix}
        1 & 0 & 0 & 0 \\
        0 & 0 & 1 & 0
    \end{bmatrix}.
\end{equation*}
The measurement noise covariance is defined as $\Sigma_v = \mathrm{diag}(\varphi, \psi)$, with $\varphi = 0.1$ and $\psi = 0.5$, corresponding to radial and angular measurement uncertainties, respectively.

To quantitatively evaluate the estimation performance, the mean square estimation error (MSEE) is computed. Let $\beta_k = \|x_k - \hat{x}_k\|$ denote the estimation error of the classical Kalman filter, and $\gamma_k = \|x_k - \hat{x}_k^{\mu}\|$ denote the estimation error of the $\mu$KF at time step $k$.
For each simulation run $j$, the MSEE is defined as
\begin{equation}
    \kappa_j = \frac{1}{N} \sum_{k=1}^{N} \beta_k^2, 
    \qquad
    \Gamma_j = \frac{1}{N} \sum_{k=1}^{N} \gamma_k^2,
    \label{eq:msee}
\end{equation}
where $N$ is the total number of time steps.
To reduce statistical variability, the simulations are repeated $\phi$ times, and the averaged MSEE is computed as
\begin{equation}
    \Xi_{\kappa} = \frac{1}{\phi} \sum_{j=1}^{\phi} \kappa_j,
    \qquad
    \Xi_{\Gamma} = \frac{1}{\phi} \sum_{j=1}^{\phi} \Gamma_j.
    \label{eq:avg_msee}
\end{equation}

\begin{figure*}[t!]
\centering
\subfloat[]{%
    \includegraphics[width=0.30\linewidth]{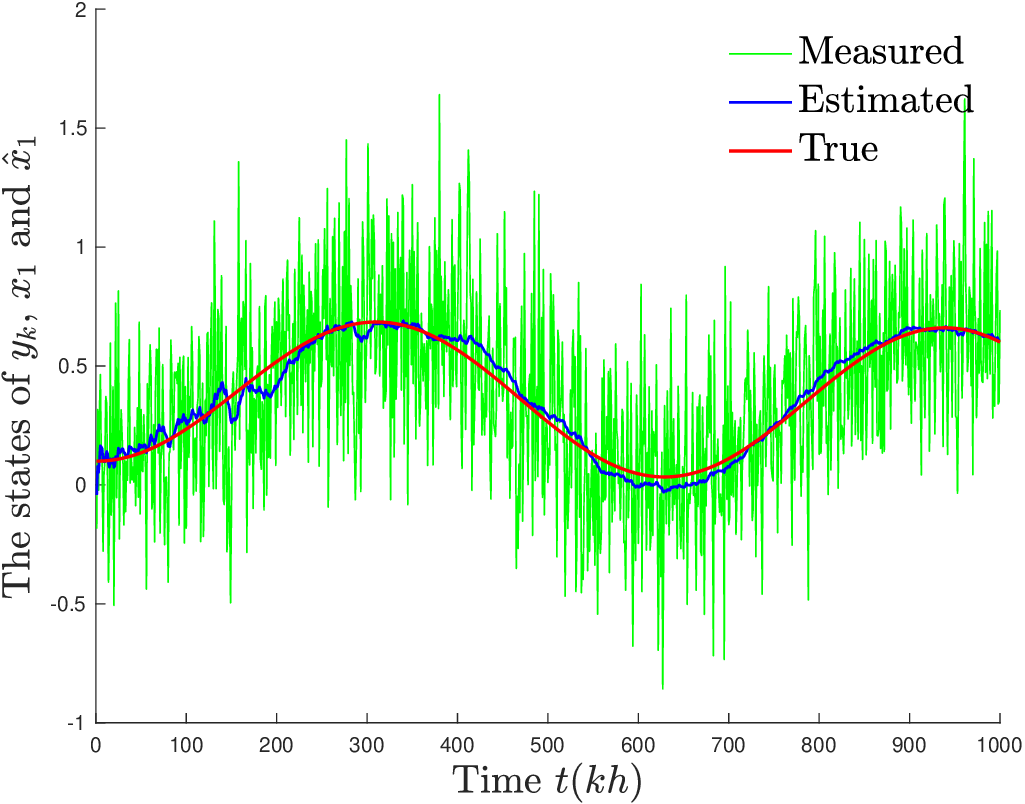}
    \label{fig:a1}
}
\subfloat[]{%
    \includegraphics[width=0.30\linewidth]{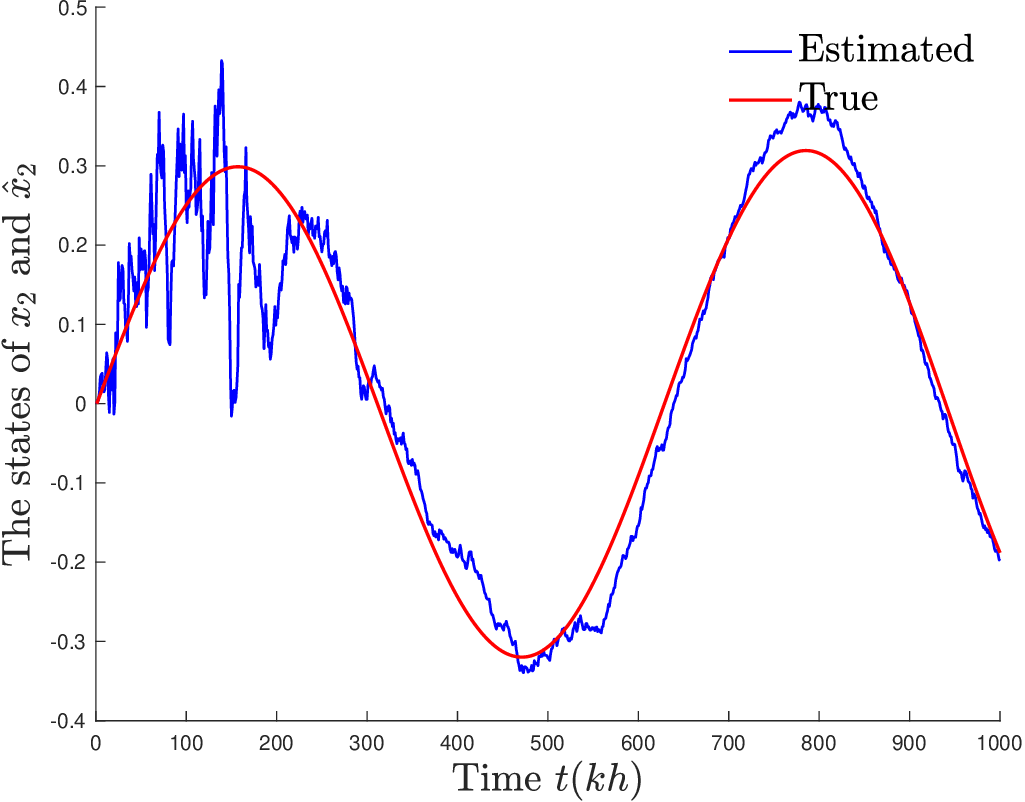}
    \label{fig:b1}
}
\subfloat[]{%
    \includegraphics[width=0.295\linewidth]{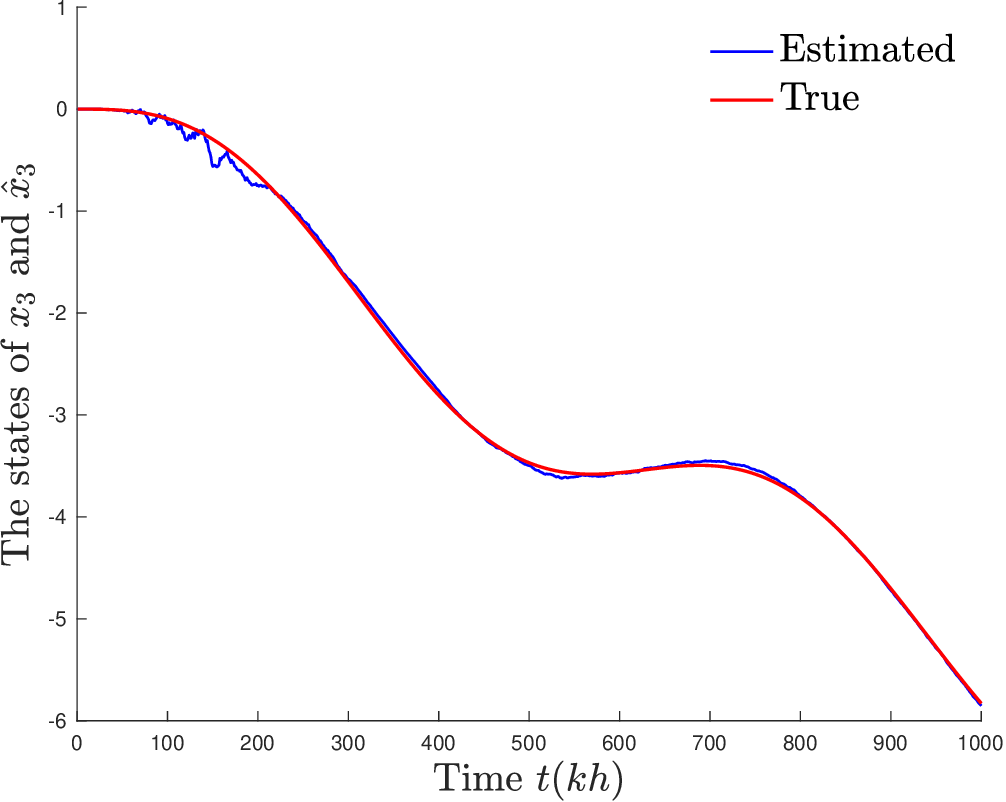}
    \label{fig:c1}
}

\medskip

\subfloat[]{%
    \includegraphics[width=0.30\linewidth]{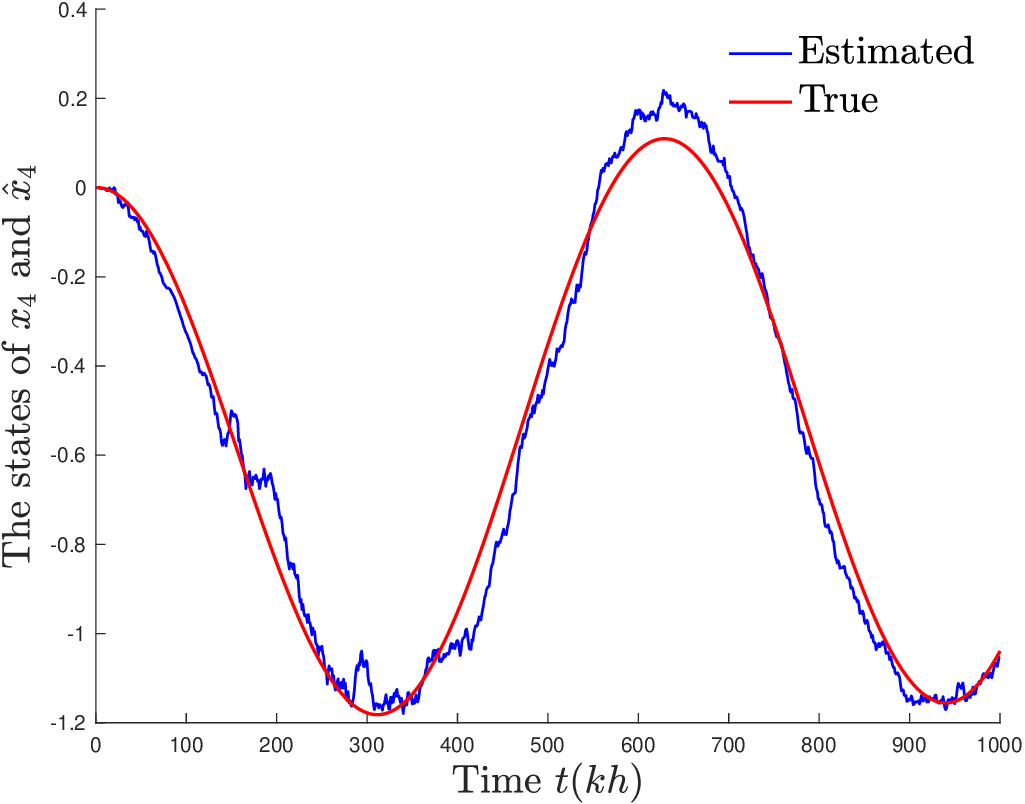}
    \label{fig:d1}
}
\subfloat[]{%
    \includegraphics[width=0.325\linewidth]{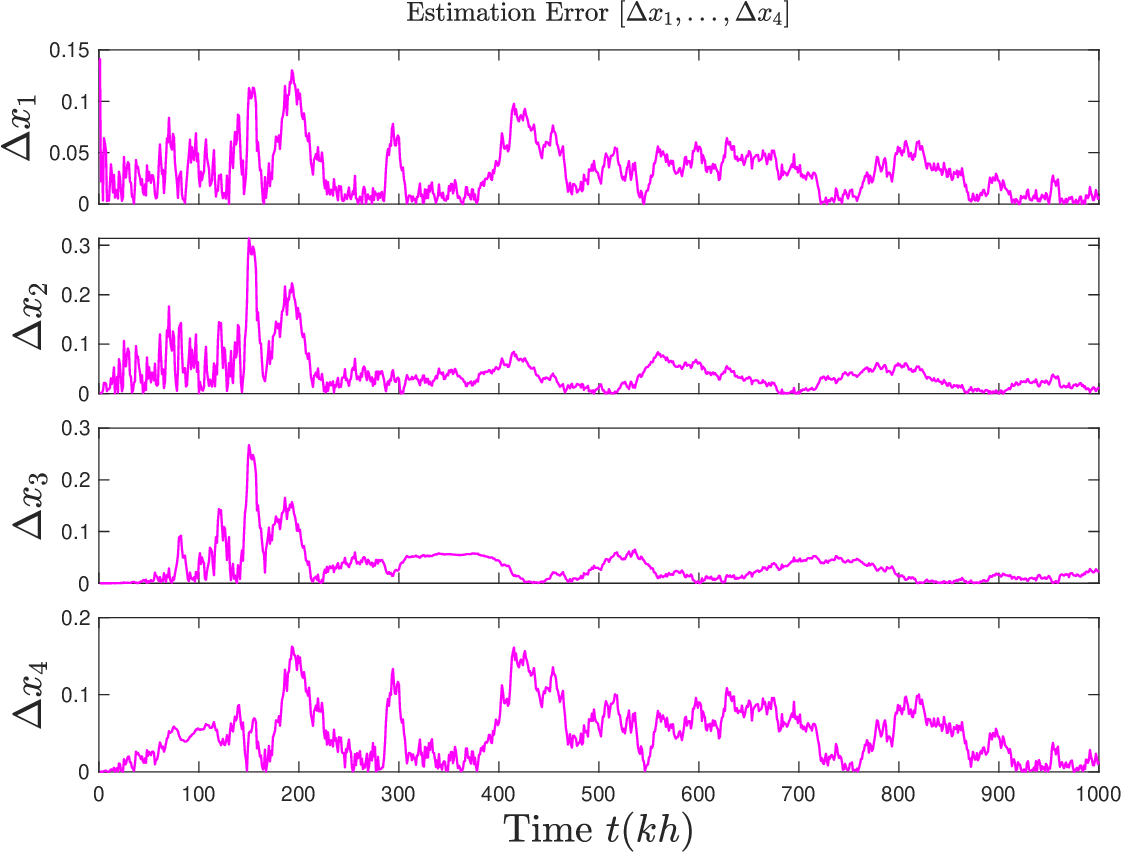}
    \label{fig:e1}
}
\caption{These figures present the results of the filtering according to four different states of position (a), velocity (b), angle (c), and angular velocity (d), whereas the last figure (e) portrays the estimation error between the true and the estimated states. The results correspond to measurements affected by noise type 1, with subplot (a) including additional measurement information.}
\label{Fig:Total1}
\end{figure*}

Figure~\ref{Fig:Total1} illustrates the true and estimated states obtained using the $\mu$KF under measurement noise type 1. The estimation errors for each state are summarized in Table~\ref{Table1}, while the corresponding error trajectories are shown in Fig.~\ref{fig:e1}. The results indicate that the $\mu$KF is able to accurately track the system states with small deviations.

A comparison between the $\mu$KF and the classical Kalman filter is presented in Table~\ref{Table3}. For noise type 1, the difference between the two methods is negligible, on the order of $10^{-4}$ across all states.

\begin{figure*}[t!]
\centering
\subfloat[]{%
    \includegraphics[width=0.30\linewidth]{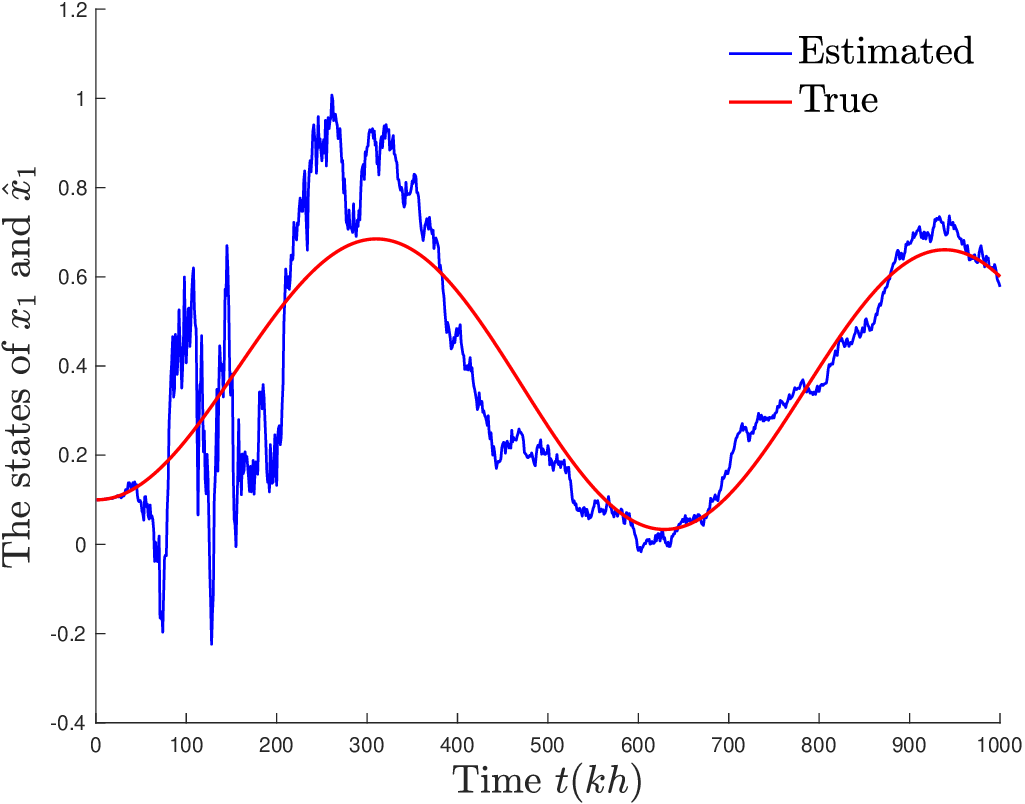}
    \label{fig:a2}
}
\subfloat[]{%
    \includegraphics[width=0.30\linewidth]{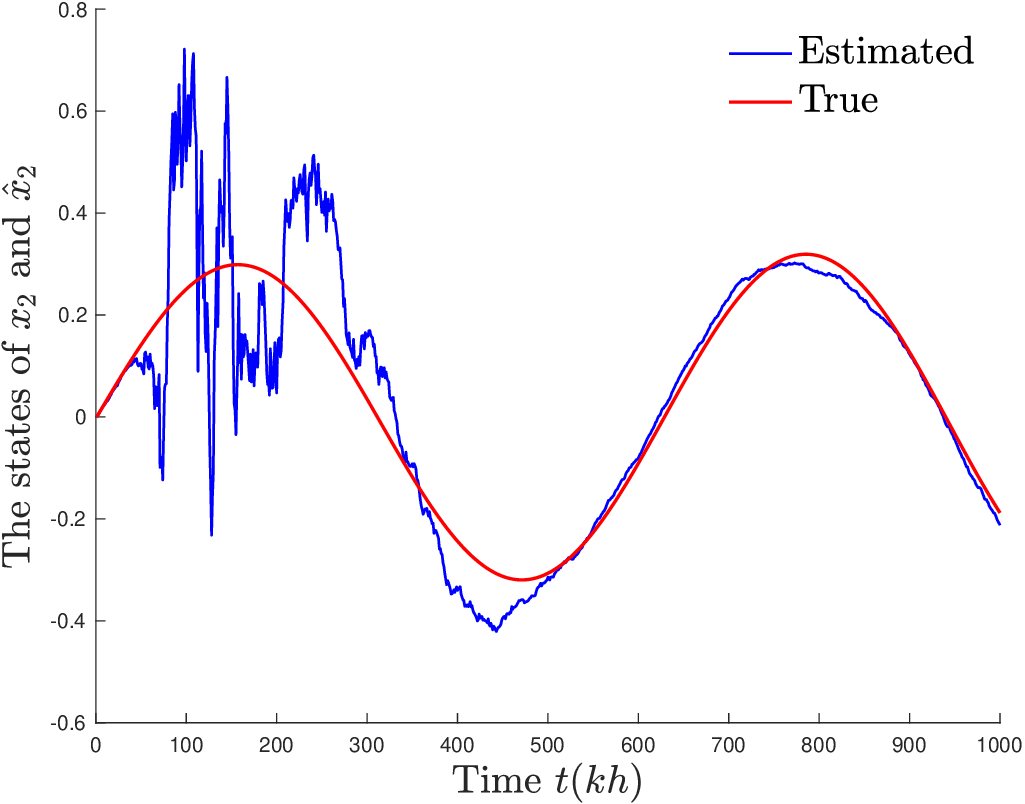}
    \label{fig:b2}
}
\subfloat[]{%
    \includegraphics[width=0.295\linewidth]{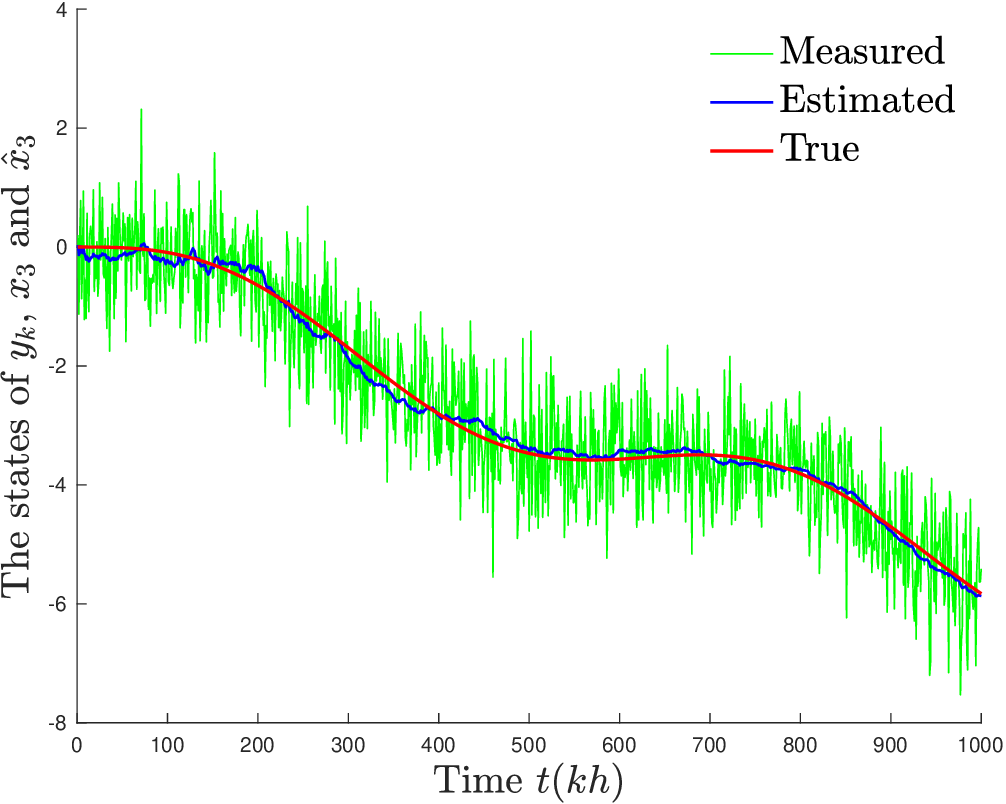}
    \label{fig:c2}
}

\medskip

\subfloat[]{%
    \includegraphics[width=0.30\linewidth]{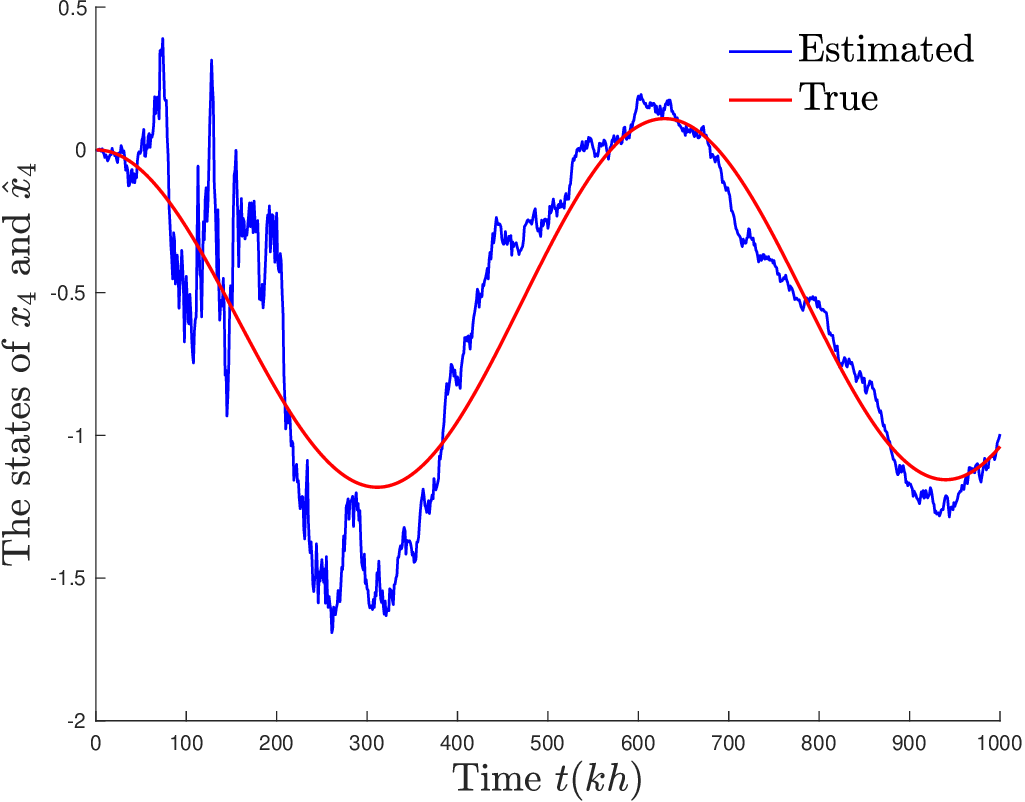}
    \label{fig:d2}
}
\subfloat[]{%
    \includegraphics[width=0.325\linewidth]{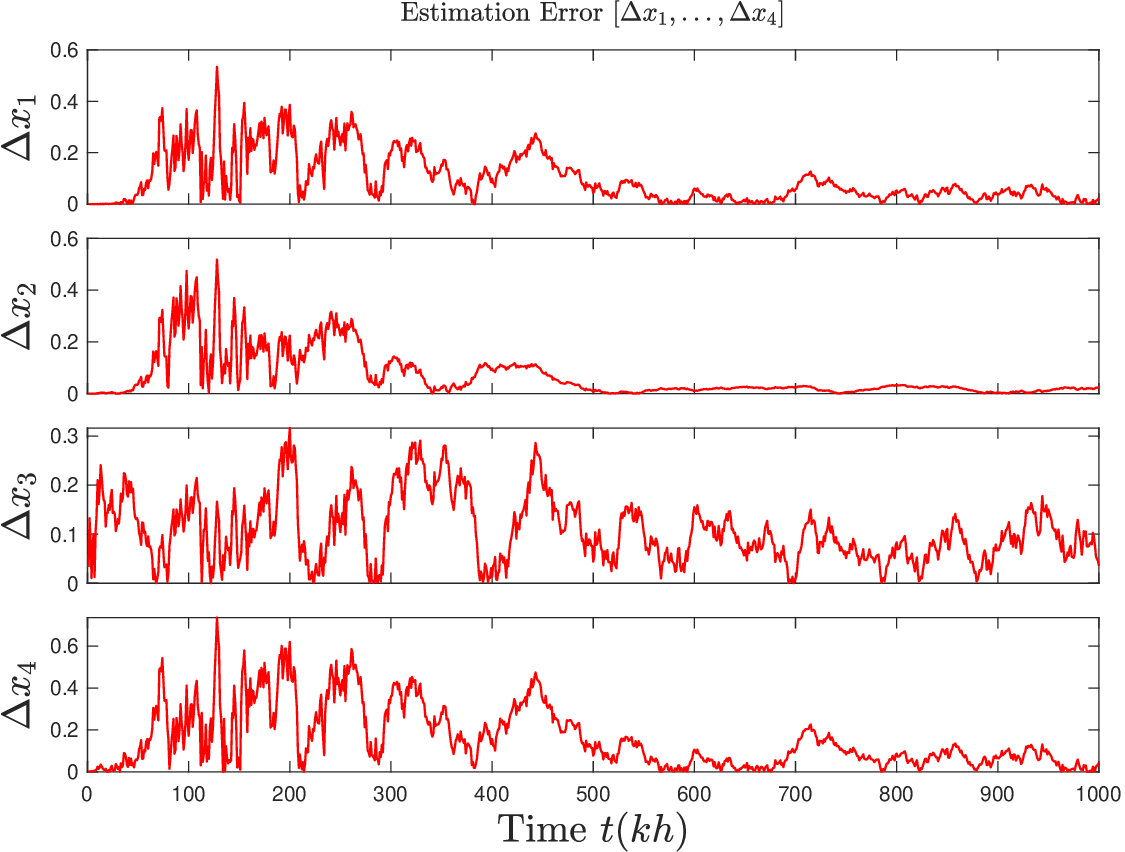}
    \label{fig:e2}
}
\caption{The figures present the filtering results for four state variables: position (a), velocity (b), angle (c), and angular velocity (d), while (e) shows the estimation error between the true and estimated states. The results correspond to measurements affected by noise type 2, with subplot (c) including additional measurement information.}
\label{Fig:Total2}
\end{figure*}

Figure~\ref{Fig:Total2} shows the corresponding results under measurement noise type 2, where the angular noise variance is increased to $\psi = 0.5$. The estimation errors increase accordingly, as shown in Table~\ref{Table2}, and the error trajectories in Fig.~\ref{fig:e2} exhibit a larger spread due to the higher measurement uncertainty.

Despite the increased noise level, the overall performance of the $\mu$KF remains comparable to that of the classical Kalman filter. As summarized in Table~\ref{Table3}, the differences between the two methods remain small, indicating that the $\mu$KF provides a consistent and reliable estimation performance under both noise configurations.

To further evaluate the proposed $\mu$KF, a comparison is conducted with the classical Kalman filter (KF), the extended Kalman filter (EKF), the unscented Kalman filter (UKF), and an adaptive Kalman filter. All methods are implemented under the same system model and noise conditions, using the combined measurement configuration defined by the observation matrix $H$.

Figure~X illustrates the estimated trajectories of the four states. It can be observed that all methods are able to track the true states with high accuracy. In particular, the $\mu$KF produces estimates that are nearly indistinguishable from those of the classical KF, confirming that both formulations are equivalent under linear Gaussian assumptions.

The EKF and UKF exhibit similar performance, as expected, since the system dynamics are linear and do not require nonlinear approximation. Consequently, their behavior closely matches that of the KF. The adaptive Kalman filter also demonstrates comparable tracking performance, with slight variations due to the online adjustment of the measurement noise covariance.

A more detailed comparison is provided in Fig.~Y, where the estimation errors are shown for each state. The results indicate that the error magnitudes of all methods remain bounded and small throughout the simulation. The $\mu$KF maintains error levels consistent with the classical KF, highlighting that the proposed formulation preserves estimation accuracy while offering an alternative computational structure.

Overall, the comparison confirms that, for linear Gaussian systems, the $\mu$KF achieves performance equivalent to the classical Kalman filter and its variants, while retaining the advantages of the information-based formulation.

\begin{table*}[t!]
\captionsetup{font=small}
\caption{The Mean Square Estimation Error (MSEE) $\Gamma_k$ with noise \textbf{type 1} $v_1$}
\centering
\resizebox{0.95\textwidth}{!}{%
\begin{tabular}{l | c c c c c c c c c c c}
\hline
\multirow{2}{*}{$x_{k+1|k}$} & \multicolumn{11}{c}{Iteration $\phi_i$} \\
& 1 & 2 & 3 & 4 & 5 & 6 & 7 & 8 & 9 & 10 & Averaged \\
\hline
$x_1$ & 0.0012 & 0.0015 & 0.0010 & 0.0043 & 0.0019 & 0.0015 & 0.0014 & 0.0011 & 0.0016 & 0.0012 & \textbf{0.0017} \\
$x_2$ & 0.0023 & 0.0023 & 0.0038 & 0.0043 & 0.0076 & 0.0048 & 0.0052 & 0.0041 & 0.0070 & 0.0061 & \textbf{0.0048} \\
$x_3$ & 0.0120 & 0.0032 & 0.0040 & 0.1271 & 0.0101 & 0.0059 & 0.0088 & 0.0039 & 0.0082 & 0.0081 & \textbf{0.0191} \\
$x_4$ & 0.0022 & 0.0034 & 0.0019 & 0.0075 & 0.0037 & 0.0022 & 0.0042 & 0.0030 & 0.0042 & 0.0037 & \textbf{0.0036} \\
\hline
\end{tabular}%
}
\label{Table1}
\end{table*}

\begin{table*}[t!]
\captionsetup{font=small}
\caption{The Mean Square Estimation Error (MSEE) $\Gamma_k$ with noise \textbf{type 2} $v_3$}
\centering
\resizebox{0.95\textwidth}{!}{%
\begin{tabular}{l | c c c c c c c c c c c}
\hline
\multirow{2}{*}{$x_{k+1|k}$} & \multicolumn{11}{c}{Iteration $\phi_i$} \\
& 1 & 2 & 3 & 4 & 5 & 6 & 7 & 8 & 9 & 10 & Averaged \\
\hline
$x_1$ & 0.0207 & 0.0365 & 0.0095 & 0.0202 & 0.0116 & 0.0150 & 0.0085 & 0.0163 & 0.0191 & 0.0117 & \textbf{0.0169} \\
$x_2$ & 0.0232 & 0.0395 & 0.0065 & 0.0205 & 0.0114 & 0.0157 & 0.0074 & 0.0159 & 0.0208 & 0.0112 & \textbf{0.0172} \\
$x_3$ & 0.0155 & 0.0180 & 0.0129 & 0.0073 & 0.0090 & 0.0080 & 0.0081 & 0.0100 & 0.0114 & 0.0084 & \textbf{0.0109} \\
$x_4$ & 0.0514 & 0.0779 & 0.0240 & 0.0389 & 0.0251 & 0.0346 & 0.0183 & 0.0353 & 0.0419 & 0.0240 & \textbf{0.0371} \\
\hline
\end{tabular}%
}
\label{Table2}
\end{table*}

\begin{table}[t!]
	\captionsetup{font=small}
	\caption{AMSEE comparison of $\Xi_{\kappa}$ and $\Xi_{\Gamma}$ with both noises}
	\centering
	\begin{tabular}{ l | c c c c}
    \hline
    \multirow{2}{4em}{$x_{k+1|k}$} & \multicolumn{2}{c}{\textbf{type 1}} & \multicolumn{2}{c}{\textbf{type 2}}\\
    & $\Xi_{\kappa}$ & $\Xi_{\Gamma}$ & $\Xi_{\kappa}$ & $\Xi_{\Gamma}$\\
	\hline
	$x_1$ & 0.0017 & \textbf{0.0016} & 0.0169 & \textbf{0.0165}\\
	$x_2$ & 0.0048 & \textbf{0.0047} & 0.0172 & \textbf{0.0169}\\
    $x_3$ & 0.0191 & \textbf{0.0193} & 0.0109 & \textbf{0.0103}\\
    $x_4$ & 0.0036 & \textbf{0.0036} & 0.0371 & \textbf{0.0361}\\
	\hline
	\end{tabular}
    \label{Table3}
\end{table}

\begin{figure*}[t!]
\centering
\subfloat[]{%
    \includegraphics[width=0.30\linewidth]{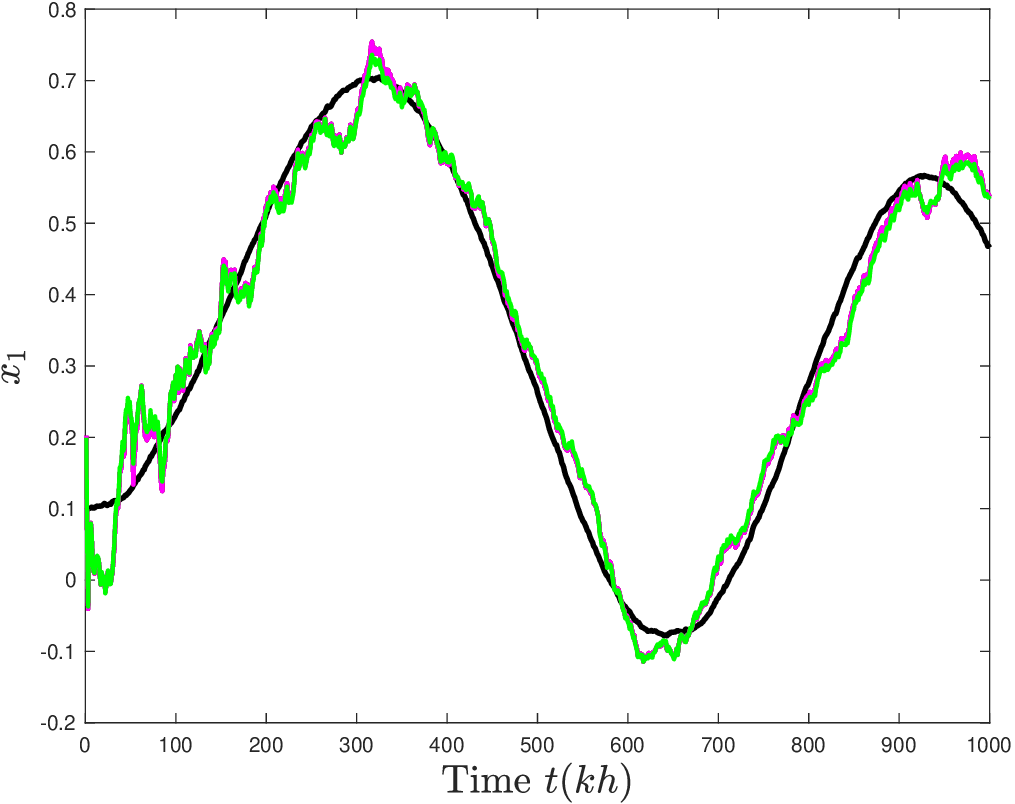}
    \label{fig:a3}
}
\subfloat[]{%
    \includegraphics[width=0.30\linewidth]{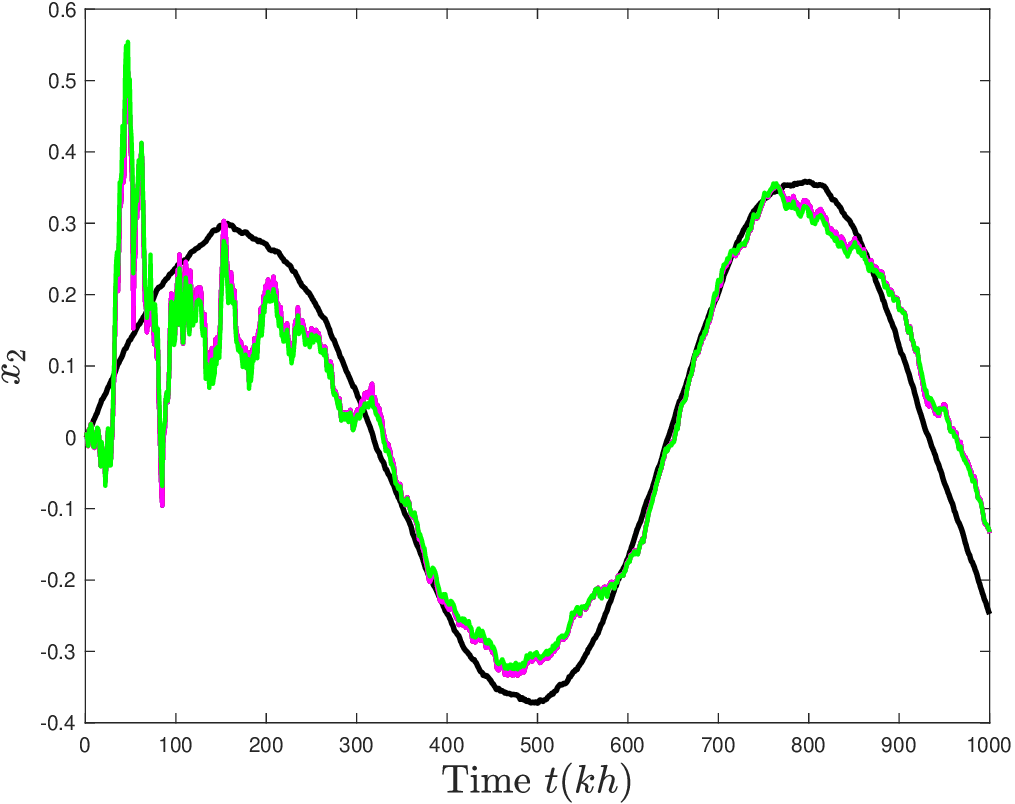}
    \label{fig:b3}
}
\subfloat[]{%
    \includegraphics[width=0.295\linewidth]{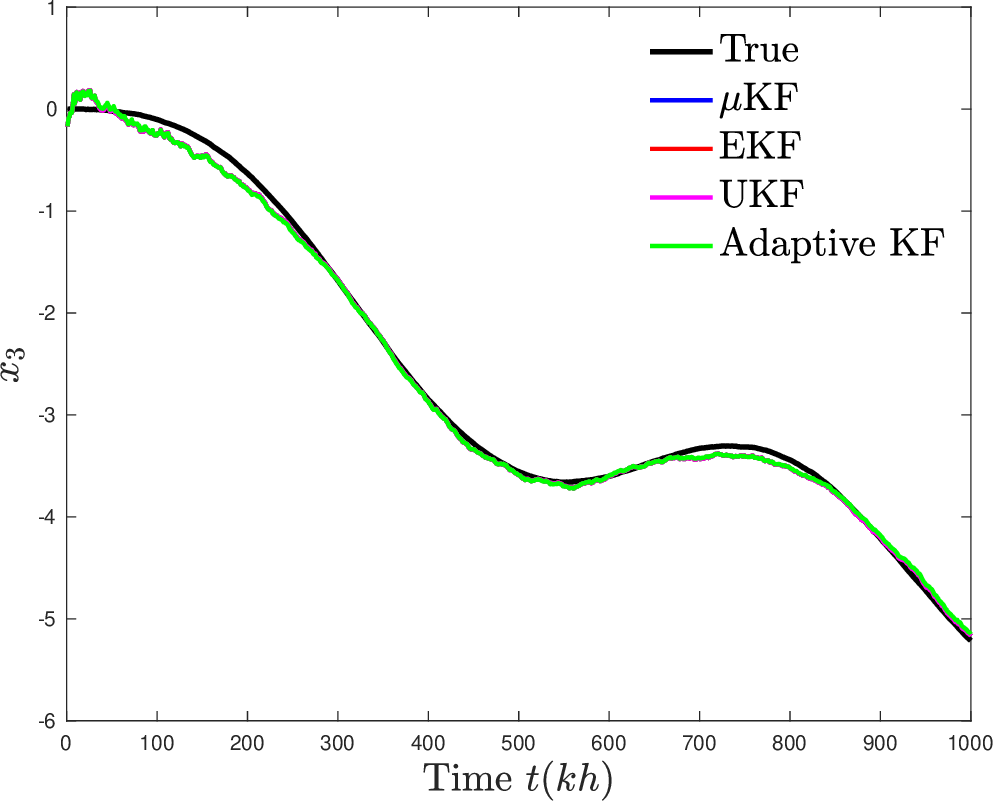}
    \label{fig:c3}
}

\medskip

\subfloat[]{%
    \includegraphics[width=0.30\linewidth]{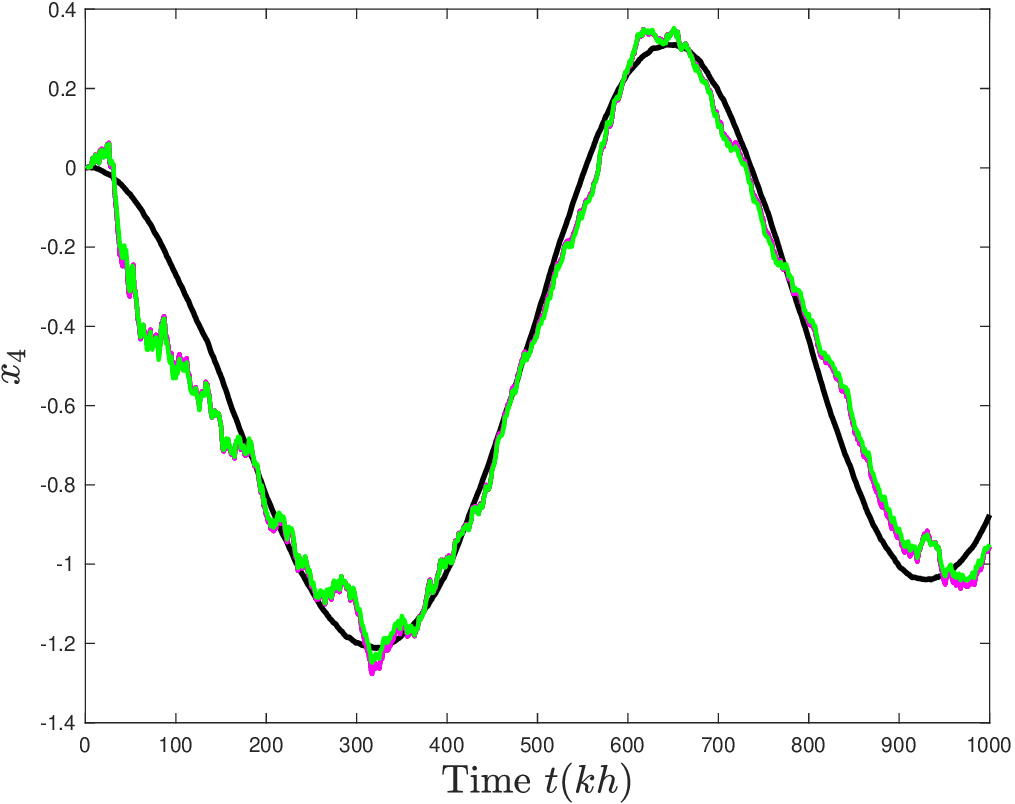}
    \label{fig:d3}
}
\subfloat[]{%
    \includegraphics[width=0.33\linewidth]{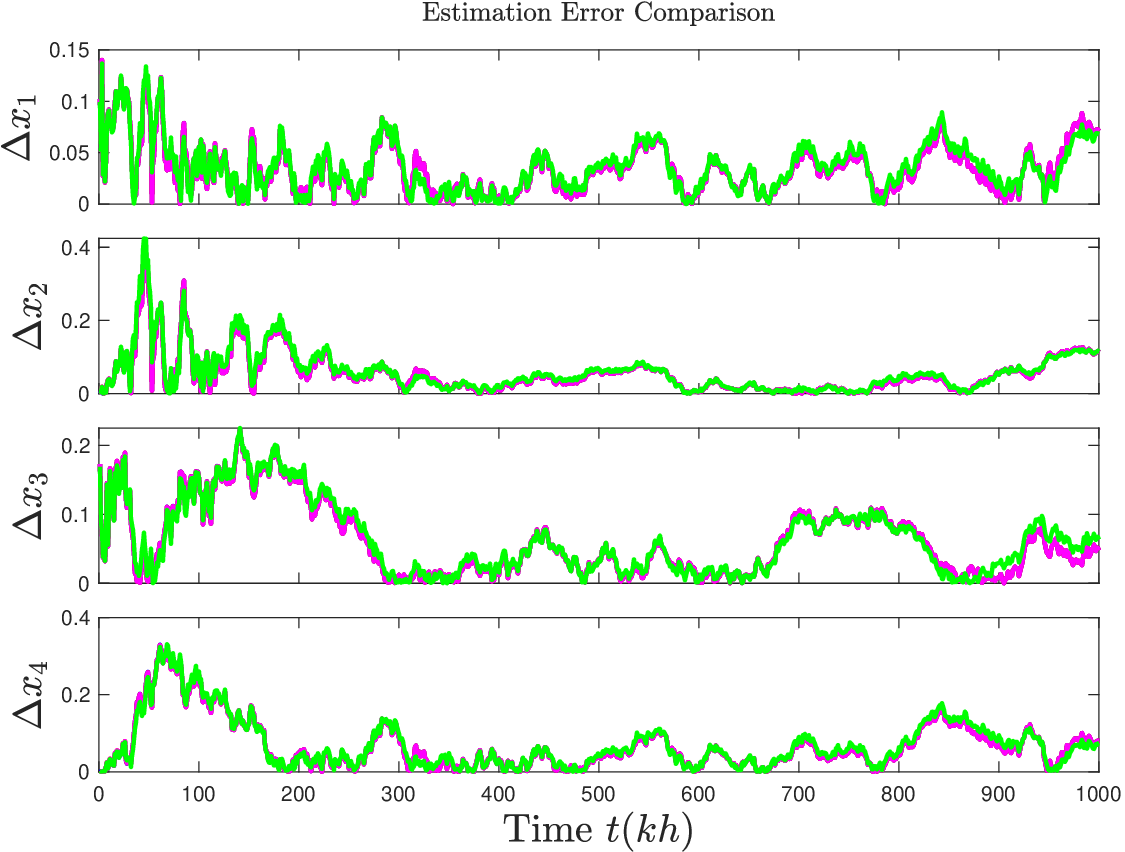}
    \label{fig:e3}
}
\caption{Comparison of filtering results for position (a), velocity (b), angle (c), and angular velocity (d) using the classical Kalman filter, $\mu$KF, EKF, UKF, and adaptive Kalman filter. The true trajectories are shown together with the corresponding estimates, while (e) presents the estimation errors. Results correspond to noise type 2, with subplot (c) including additional measurement information.}
\label{Fig:Total3}
\end{figure*}

\section{Conclusion}

This paper presented the mathematical formulation of the classical Kalman filter and the proposed $\mu$KF within a unified state-space framework, together with the construction of two types of noisy measurements. The performance of the $\mu$KF was evaluated through numerical simulations under different noise configurations.

The results demonstrate that the $\mu$KF is capable of accurately tracking the satellite dynamics, with mean square estimation errors (MSEE) reported in Table~\ref{Table1} and Table~\ref{Table2}. Furthermore, the comparison in Table~\ref{Table3} shows that the estimation performance of the $\mu$KF is nearly identical to that of the classical Kalman filter.

A comparative study with the extended Kalman filter (EKF), unscented Kalman filter (UKF), and adaptive Kalman filter further confirms that all methods achieve similar estimation accuracy under the considered linear Gaussian setting. In particular, the $\mu$KF produces results that are consistent with these approaches, while maintaining a formulation based on information matrices.

These findings confirm that the $\mu$KF preserves the estimation accuracy of the standard Kalman filter and its variants, while offering an alternative computational structure. This makes the proposed approach a viable and effective method for state estimation in linear Gaussian systems.

\section{Future Works}

The present work focuses on state estimation for linear Gaussian systems using Kalman-based approaches. While the results demonstrate that the proposed $\mu$KF achieves performance equivalent to the classical Kalman filter, several research directions can be pursued to extend this framework toward more complex and realistic scenarios.

A natural extension is the application of the proposed estimation framework to nonlinear systems. In such settings, techniques such as exact linearization and unscented filtering become essential to capture the system dynamics accurately. For instance, adaptive Kalman filtering combined with exact linearization and decoupling control has shown effectiveness in process systems such as multi-tank configurations, where strong nonlinearities are present \cite{Wafi-ThreeTank,F1a,F1b}. In addition, consensus-based unscented filtering provides a promising approach for nonlinear estimation in vehicle dynamics, particularly in autonomous and sensorless designs \cite{WAFI-JRC,F2a,F2b}.

Another important direction is the incorporation of adaptive mechanisms for handling uncertainties and disturbances. In practical systems, noise statistics are often unknown or time-varying. Adaptive Kalman filtering strategies, including covariance adaptation and scaling techniques, can improve robustness and estimation accuracy. These approaches are particularly relevant in applications such as hydraulic systems, where fault detection and isolation are critical for system reliability \cite{Wafi-Hydraulic,F3a,F3b}.

The framework can also be extended to distributed estimation in networked systems. In multi-agent systems, decentralized and consensus-based filtering methods enable each agent to estimate the global state using local information and limited communication. This is especially important for large-scale interconnected processes, such as quadruple-tank systems, where distributed estimation must be integrated with decentralized control strategies \cite{Wafi-Quadruple,F4a,F4b}.

Finally, a significant research direction lies in the integration of estimation and control. Combining state estimation with advanced control techniques, such as model reference adaptive control and robust non-fragile control, allows the development of closed-loop systems capable of handling uncertainties, disturbances, and delays \cite{Wafi-Elham,F5a,F5b}. In particular, distributed adaptive control for interconnected systems and networked control with state and input delays represent challenging yet impactful extensions of the present work \cite{Wafi-LCSS24,F6a,F6b,Wafi-MRAC,F7a,F7b}.

Overall, these directions provide a pathway from centralized linear estimation toward nonlinear, distributed, and control-integrated frameworks, enabling the application of the proposed methods to a broader class of real-world systems.

\section*{Acknowledgment}
Thanks to Professor Richard B. Vinter from the Imperial College London who has taught me in the lecture leading to finishing this paper and to LPDP (Indonesia Endowment Fund for Education) Scholarship from Indonesia.

\bibliographystyle{ieeetr}
\bibliography{Ref.bib}

@article{b1,
  author  = {F. A. Faruqi and R. C. Davis},
  title   = {Kalman Filter Design for Target Tracking},
  journal = {IEEE Transactions on Aerospace and Electronic Systems},
  volume  = {AES-16},
  number  = {4},
  pages   = {},
  month   = {July},
  year    = {1980}
}

@article{b2,
  author  = {B. Esmat},
  title   = {An Adaptive Kalman Filter for Tracking Maneuvering Targets},
  journal = {Journal of Guidance, Control, and Dynamics},
  volume  = {6},
  number  = {5},
  pages   = {414--416},
  month   = {September},
  year    = {1983}
}

@article{b3,
  author  = {W. Yuh-Shyang and S. You and N. Matni},
  title   = {Localized Distributed Kalman Filters for Large-Scale Systems},
  journal = {IFAC-PapersOnLine},
  volume  = {50},
  number  = {1},
  pages   = {10742--10747},
  year    = {2017},
  publisher = {Elsevier}
}

@article{b4,
  author  = {R. E. Kalman},
  title   = {A New Approach to Linear Filtering and Prediction Problems},
  journal = {ASME Journal of Basic Engineering},
  year    = {1960}
}

@book{b5,
  author    = {B. D. O. Anderson and J. B. Moore},
  title     = {Optimal Filtering},
  publisher = {Dover},
  address   = {New York},
  year      = {2005}
}

@article{b6,
  author  = {E. Mansour},
  title   = {Book Review: Optimal Filtering},
  journal = {IEEE Transactions on Systems, Man, and Cybernetics},
  volume  = {SMC-12},
  number  = {2},
  pages   = {},
  month   = {March/April},
  year    = {1982}
}

@inproceedings{b7,
  author    = {R. Olfati-Saber},
  title     = {Distributed Kalman Filter with Embedded Consensus Kalman Filters},
  booktitle = {Proceedings of the 44th IEEE Conference on Decision and Control and the European Control Conference},
  address   = {Seville, Spain},
  month     = {December},
  year      = {2005}
}

@inproceedings{b8,
  author    = {R. Olfati-Saber},
  title     = {Distributed Kalman Filtering for Sensor Networks},
  booktitle = {Proceedings of the 46th IEEE Conference on Decision and Control},
  address   = {New Orleans, LA, USA},
  month     = {December},
  year      = {2007}
}

@article{b9,
  author  = {M. B. Rhudy and R. A. Salguero and K. Holappa},
  title   = {A Kalman Filtering Tutorial for Undergraduate Students},
  journal = {International Journal of Computer Science and Engineering Survey (IJCSES)},
  volume  = {8},
  number  = {1},
  month   = {February},
  year    = {2017}
}

@ARTICLE{F1a,
  author={Ge, Quanbo and Hu, Xiaoming and Li, Yunyu and He, Hongli and Song, Zihao},
  journal={IEEE/CAA Journal of Automatica Sinica}, 
  title={A Novel Adaptive Kalman Filter Based on Credibility Measure}, 
  year={2023},
  volume={10},
  number={1},
  pages={103-120},
  doi={10.1109/JAS.2023.123012}
}

@ARTICLE{F1b,
  author={Kruse, Theresa and Griebel, Thomas and Graichen, Knut},
  journal={IEEE Access}, 
  title={Adaptive Kalman Filtering: Measurement and Process Noise Covariance Estimation Using Kalman Smoothing}, 
  year={2025},
  volume={13},
  number={},
  pages={11863-11875},
  doi={10.1109/ACCESS.2025.3528348}
}

@ARTICLE{F2a,
  author={Kappl, J. J.},
  journal={IEEE Transactions on Aerospace and Electronic Systems}, 
  title={Nonlinear Estimation Via Kalman Filtering}, 
  year={1971},
  volume={AES-7},
  number={1},
  pages={79-84},
  doi={10.1109/TAES.1971.310255}}

@INPROCEEDINGS{F2b,
  author={Schumann, Markus and Ebersberger, Sebastian and Graichen, Knut},
  booktitle={2023 IEEE International Conference on Mechatronics (ICM)}, 
  title={Improved Nonlinear Estimation in Thermal Networks Using Machine Learning}, 
  year={2023},
  volume={},
  number={},
  pages={1-6},
  doi={10.1109/ICM54990.2023.10102071}}

@INPROCEEDINGS{F3a,
  author={Barzegar, Ailin and Rahimi, Afshin},
  booktitle={2023 IEEE International Conference on Prognostics and Health Management (ICPHM)}, 
  title={A Distributed Fault Detection and Estimation for Formation of Clusters of Small Satellites}, 
  year={2023},
  volume={},
  number={},
  pages={1-11},
  doi={10.1109/ICPHM57936.2023.10194041}}

@INPROCEEDINGS{F3b,
  author={Alessandri, Angelo and Boem, Francesca and Parisini, Thomas},
  booktitle={2018 IEEE Conference on Decision and Control (CDC)}, 
  title={Model-Based Fault Detection and Estimation for Linear Time Invariant and Piecewise Affine Systems by Using Quadratic Boundedness}, 
  year={2018},
  volume={},
  number={},
  pages={5562-5567},
  doi={10.1109/CDC.2018.8619344}}

@INPROCEEDINGS{F4a,
  author={Sihag, Saurabh and Tajer, Ali},
  booktitle={2018 IEEE International Conference on Acoustics, Speech and Signal Processing (ICASSP)}, 
  title={Distributed Estimation Under Network Model Uncertainty}, 
  year={2018},
  volume={},
  number={},
  pages={3569-3573},
  doi={10.1109/ICASSP.2018.8461488}}

@ARTICLE{F4b,
  author={Doostmohammadian, Mohammadreza and Taghieh, Amin and Zarrabi, Houman},
  journal={IEEE Transactions on Automation Science and Engineering}, 
  title={Distributed Estimation Approach for Tracking a Mobile Target via Formation of UAVs}, 
  year={2022},
  volume={19},
  number={4},
  pages={3765-3776},
  doi={10.1109/TASE.2021.3135834}}

@ARTICLE{F5a,
  author={Mili, L. and Steeno, G. and Dobraca, F. and French, D.},
  journal={IEEE Transactions on Power Systems}, 
  title={A robust estimation method for topology error identification}, 
  year={1999},
  volume={14},
  number={4},
  pages={1469-1476},
  doi={10.1109/59.801932}}

@ARTICLE{F5b,
  author={Zoubir, Abdelhak M. and Koivunen, Visa and Chakhchoukh, Yacine and Muma, Michael},
  journal={IEEE Signal Processing Magazine}, 
  title={Robust Estimation in Signal Processing: A Tutorial-Style Treatment of Fundamental Concepts}, 
  year={2012},
  volume={29},
  number={4},
  pages={61-80},
  doi={10.1109/MSP.2012.2183773}}

@ARTICLE{F6a,
  author={Pal, Diptak and Panigrahi, Bijaya Ketan and Bhasin, Shubhendu},
  journal={IEEE Systems Journal}, 
  title={Distributed Adaptive Control Framework for Enhanced Voltage and Frequency Regulation in Inverter Interfaced Autonomous Distribution Network}, 
  year={2023},
  volume={17},
  number={2},
  pages={2892-2903},
  doi={10.1109/JSYST.2022.3215760}}

@INPROCEEDINGS{F6b,
  author={Syed, Wasif H. and Machado, Juan E. and Schiffer, Johannes},
  booktitle={2024 American Control Conference (ACC)}, 
  title={Distributed Adaptive Control for a DC Power Distribution System of a Series-Hybrid-Electric Propulsion System of a Commuter Aircraft}, 
  year={2024},
  volume={},
  number={},
  pages={2598-2603},
  doi={10.23919/ACC60939.2024.10644618}}

@INPROCEEDINGS{F7a,
  author={Renganathan, Venkatraman and Rantzer, Anders and Kjellqvist, Olle},
  booktitle={2024 European Control Conference (ECC)}, 
  title={Distributed Adaptive Control For Uncertain Networks}, 
  year={2024},
  volume={},
  number={},
  pages={1789-1794},
  doi={10.23919/ECC64448.2024.10591151}}

@ARTICLE{F7b,
  author={Yu, Wenwu and DeLellis, Pietro and Chen, Guanrong and di Bernardo, Mario and Kurths, Jürgen},
  journal={IEEE Transactions on Automatic Control}, 
  title={Distributed Adaptive Control of Synchronization in Complex Networks}, 
  year={2012},
  volume={57},
  number={8},
  pages={2153-2158},
  doi={10.1109/TAC.2012.2183190}}

@article{Wafi-MRAC,
   title={Model reference adaptive control of networked systems with state and input delays},
   volume={14},
   ISSN={2088-8708},
   DOI={10.11591/ijece.v14i5.pp5055-5063},
   number={5},
   journal={International Journal of Electrical and Computer Engineering (IJECE)},
   publisher={Institute of Advanced Engineering and Science},
   author={Wafi, Moh Kamalul and Indriawati, Katherin and Widjiantoro, Bambang L.},
   year={2024},
   month=oct, 
   pages={5055} 
}

@article{Wafi-Elham,
author = {Javanfar, Elham and Rahmani, Mehdi and Wafi, Moh Kamalul},
title = {Robust Estimation-Based Non-Fragile Control for Discrete-Time Non-Linear Systems},
journal = {International Journal of Robust and Nonlinear Control},
volume = {35},
number = {6},
pages = {2462-2471},
eprint = {https://onlinelibrary.wiley.com/doi/pdf/10.1002/rnc.7806},
year = {2025}
}

@article{WAFI-JRC, 
title={Non-Linear Estimation using the Weighted Average Consensus-Based Unscented Filtering for Various Vehicles Dynamics towards Autonomous Sensorless Design}, 
volume={4}, 
number={1}, 
journal={Journal of Robotics and Control (JRC)}, 
author={Widjiantoro, Bambang L. and Wafi, Moh Kamalul and Indriawati, Katherin}, 
year={2023}, 
month={Mar.}, 
pages={95–107},
url={https://journal.umy.ac.id/index.php/jrc/article/view/16164}, 
DOI={10.18196/jrc.v4i1.16164},
}

@ARTICLE{Wafi-LCSS24,
  author={Wafi, Moh. Kamalul and Siami, Milad},
  journal={IEEE Control Systems Letters}, 
  title={Distributed Adaptive Control of Disturbed Interconnected Systems With High-Order Tuners}, 
  year={2024},
  volume={8},
  number={},
  pages={1421-1426},
  doi={10.1109/LCSYS.2024.3407613}
}

@article{Wafi-Hydraulic,
    title={Estimation and Fault Detection on Hydraulic System with Adaptive-Scaling Kalman and Consensus Filtering}, 
    author={Moh Kamalul Wafi},
    journal = {arXiv preprint arXiv:2304.04122},
    year = {2023},
}

@article{Wafi-Quadruple,
    title={Distributed Estimation with Decentralized Control for Quadruple-Tank Process}, 
    author={Moh Kamalul Wafi and Bambang L. Widjiantoro},
    journal = {arXiv preprint arXiv:2304.04763},
    year = {2025},
}

@article{Wafi-ThreeTank,
    title={Adaptive Kalman Filtering with Exact Linearization and Decoupling Control on Three-Tank Process}, 
    author={Bambang L. Widjiantoro and Katherin Indriawati and Moh Kamalul Wafi},
    journal = {arXiv preprint arXiv:2304.04144},
    year = {2026},
}

\appendix
\section{Appendix}
\noindent\textbf{System Model:}
\begin{align}
    \begin{aligned}
    x_k &= F x_{k-1} + q_{k-1} \\
    y_k &= H x_k + v_k
    \end{aligned}
\end{align}
with $q_k \sim \mathcal{N}(0, \Sigma_q)$ and $v_k \sim \mathcal{N}(0, \Sigma_v)$.

\begin{algorithm}
\caption{$\mu$-Kalman Filter (Information Form)}
\begin{algorithmic}[1]
\STATE Initialize $\hat{x}_{0|0}, P_{0|0}$
\FOR{$k = 1,2,\dots,N$}
    \STATE \textbf{Prediction:}
    \begin{align*}
        \hat{x}_{k|k-1} &= F \hat{x}_{k-1|k-1} \\
        P_{k|k-1} &= F P_{k-1|k-1} F^\top + \Sigma_q
    \end{align*}
    \STATE \textbf{Information Update:}
    \begin{align*}
        S_k &= H^\top \Sigma_v^{-1} H \\
        M_k &= \left(P_{k|k-1}^{-1} + S_k \right)^{-1} \\
        z_k &= H^\top \Sigma_v^{-1} y_k
    \end{align*}
    \STATE \textbf{State Update:}
    \begin{align*}
        \hat{x}_{k|k} = \hat{x}_{k|k-1} + M_k (z_k - S_k \hat{x}_{k|k-1})
    \end{align*}
\ENDFOR
\end{algorithmic}
\end{algorithm}

\begin{algorithm}
\caption{Extended Kalman Filter (EKF)}
\begin{algorithmic}[1]
\STATE Initialize $\hat{x}_{0|0}, P_{0|0}$
\FOR{$k = 1,2,\dots,N$}
    \STATE \textbf{Prediction:}
    \begin{align*}
        \hat{x}_{k|k-1} &= f(\hat{x}_{k-1|k-1}), \qquad 
        F_k = \left.\frac{\partial f}{\partial x}\right|_{\hat{x}_{k-1|k-1}} \\
        P_{k|k-1} &= F_k P_{k-1|k-1} F_k^\top + \Sigma_q
    \end{align*}
    \STATE \textbf{Update:}
    \begin{align*}
        H_k &= \left.\frac{\partial h}{\partial x}\right|_{\hat{x}_{k|k-1}} \\
        K_k &= P_{k|k-1} H_k^\top (H_k P_{k|k-1} H_k^\top + \Sigma_v)^{-1} \\
        \hat{x}_{k|k} &= \hat{x}_{k|k-1} + K_k (y_k - h(\hat{x}_{k|k-1})) \\
        P_{k|k} &= (I - K_k H_k) P_{k|k-1}
    \end{align*}
\ENDFOR
\end{algorithmic}
\end{algorithm}

\begin{algorithm}
\caption{Unscented Kalman Filter (UKF)}
\begin{algorithmic}[1]
\STATE Initialize $\hat{x}_{0|0}, P_{0|0}$
\FOR{$k = 1,2,\dots,N$}
    \STATE \textbf{Sigma Points:}
    \begin{align*}
        \chi_{k-1}^{(i)} = \text{SigmaPoints}(\hat{x}_{k-1|k-1}, P_{k-1|k-1})
    \end{align*}
    \STATE \textbf{Prediction:}
    \begin{align*}
        \chi_k^{(i)} &= f(\chi_{k-1}^{(i)}) \\
        \hat{x}_{k|k-1} &= \sum_i W_i \chi_k^{(i)} \\
        P_{k|k-1} &= \sum_i W_i (\chi_k^{(i)} - \hat{x}_{k|k-1})(\cdot)^\top + \Sigma_q
    \end{align*}
    \STATE \textbf{Measurement Prediction:}
    \begin{align*}
        y_k^{(i)} &= h(\chi_k^{(i)}) \quad \longrightarrow \quad
        \hat{y}_k = \sum_i W_i y_k^{(i)}
    \end{align*}
    \STATE \textbf{Update:}
    \begin{align*}
        P_{yy} &= \sum_i W_i (y_k^{(i)} - \hat{y}_k)(\cdot)^\top + \Sigma_v, \qquad 
        P_{xy} = \sum_i W_i (\chi_k^{(i)} - \hat{x}_{k|k-1})(y_k^{(i)} - \hat{y}_k)^\top \\
        K_k &= P_{xy} P_{yy}^{-1} \\
        \hat{x}_{k|k} &= \hat{x}_{k|k-1} + K_k (y_k - \hat{y}_k) \\
        P_{k|k} &= P_{k|k-1} - K_k P_{yy} K_k^\top
    \end{align*}
\ENDFOR
\end{algorithmic}
\end{algorithm}


\begin{algorithm}
\caption{Adaptive Kalman Filter (Covariance Estimation)}
\begin{algorithmic}[1]
\STATE Initialize $\hat{x}_{0|0}, P_{0|0}, \Sigma_q, \Sigma_v$
\FOR{$k = 1,2,\dots,N$}
    \STATE \textbf{Prediction:}
    \begin{align*}
        \hat{x}_{k|k-1} &= F \hat{x}_{k-1|k-1} \\
        P_{k|k-1} &= F P_{k-1|k-1} F^\top + \Sigma_q
    \end{align*}
    \STATE \textbf{Innovation:}
    \begin{align*}
        e_k = y_k - H \hat{x}_{k|k-1}
    \end{align*}
    \STATE \textbf{Adaptive Update:}
    \begin{align*}
        \Sigma_v &= \alpha \Sigma_v + (1-\alpha) (e_k e_k^\top)
    \end{align*}
    \STATE \textbf{Kalman Update:}
    \begin{align*}
        K_k &= P_{k|k-1} H^\top (H P_{k|k-1} H^\top + \Sigma_v)^{-1} \\
        \hat{x}_{k|k} &= \hat{x}_{k|k-1} + K_k e_k \\
        P_{k|k} &= (I - K_k H) P_{k|k-1}
    \end{align*}
\ENDFOR
\end{algorithmic}
\end{algorithm}

\begin{figure*}[t!]
\centering
\subfloat[]{%
    \includegraphics[width=0.33\linewidth]{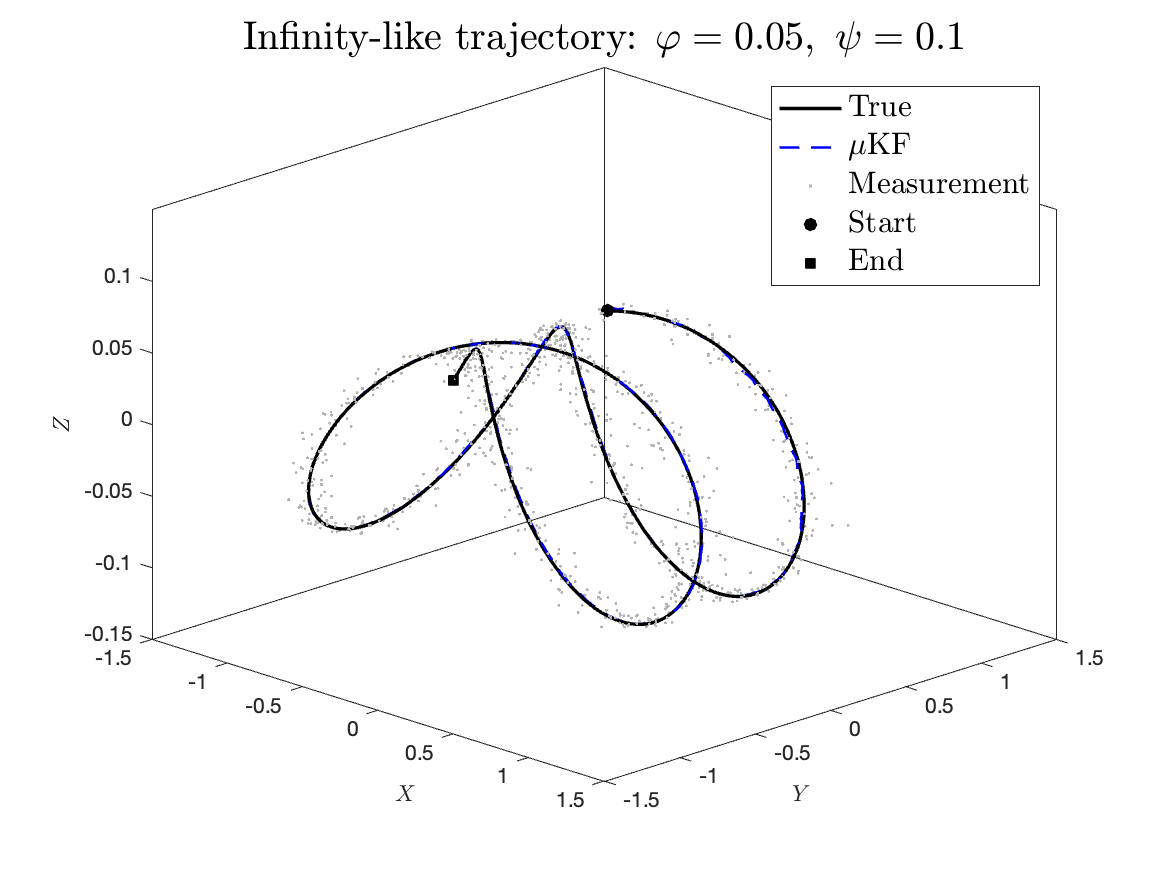}
    \label{fig:inf_a}
}
\subfloat[]{%
    \includegraphics[width=0.33\linewidth]{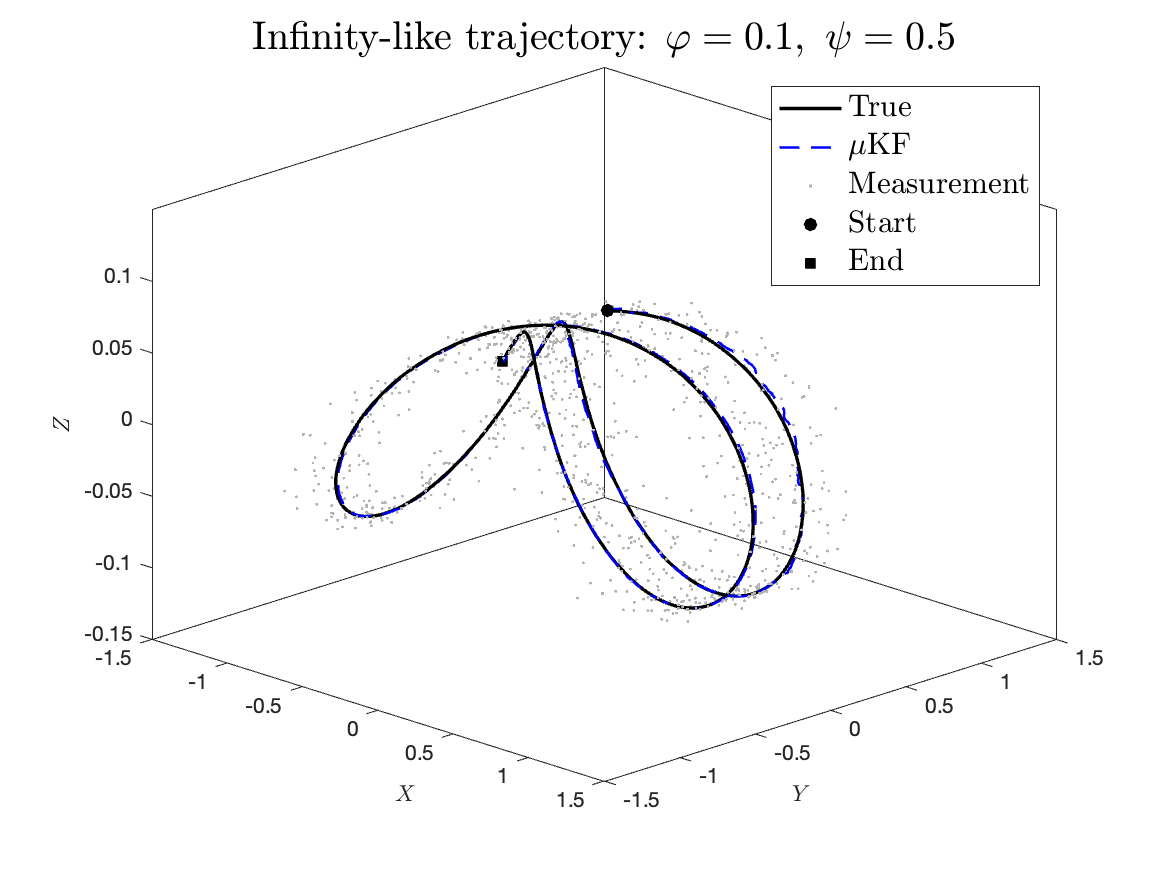}
    \label{fig:inf_b}
}
\subfloat[]{%
    \includegraphics[width=0.33\linewidth]{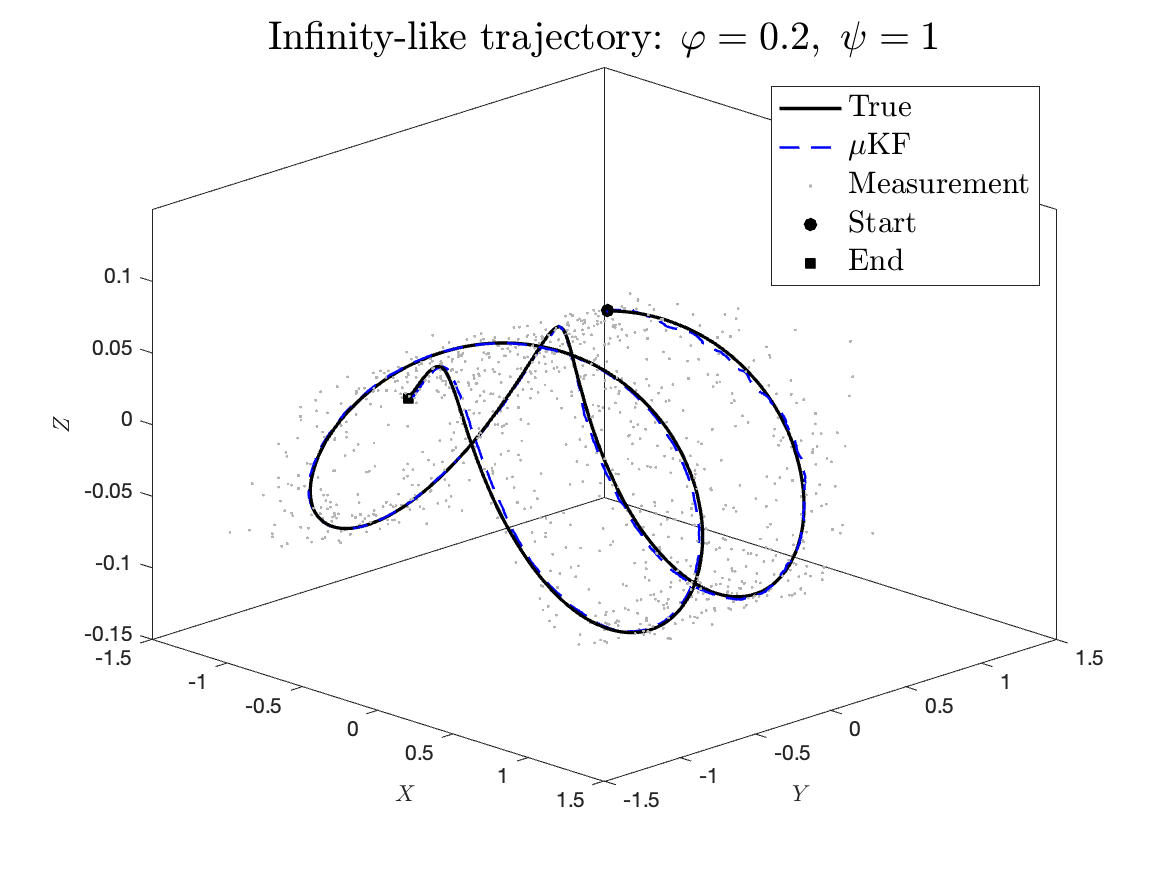}
    \label{fig:inf_c}
}
\caption{Infinity-like trajectory reconstruction under different noise configurations: (a) low noise $(\varphi = 0.05, \psi = 0.10)$, (b) moderate noise $(\varphi = 0.10, \psi = 0.50)$, and (c) high noise $(\varphi = 0.20, \psi = 1.00)$. The true trajectory, noisy measurements, and the $\mu$KF estimates are shown for comparison.}
\label{fig:infinity_compare}
\end{figure*}

\end{document}